\documentclass[11pt]{article}
\usepackage{marginnote}

\usepackage{color}
\usepackage{amsmath}
\usepackage{amsthm}
\usepackage{comment}
\usepackage{marginnote}
\allowdisplaybreaks

\usepackage{array}
\usepackage{tabularx}
\usepackage{appendix}
\usepackage{makecell}
\usepackage{graphicx}
\usepackage{amssymb}
\usepackage[T1]{fontenc}
\usepackage[utf8]{inputenc}
\usepackage[raggedright]{titlesec}
\usepackage{blindtext}
\usepackage{commath}
\usepackage{cite}
\usepackage{caption}
\usepackage[textwidth = 7in]{geometry}
\usepackage{mathtools}
\usepackage[dvipsnames]{xcolor}
\usepackage{enumitem}
\usepackage{amsmath}
\usepackage{geometry}
\usepackage{lipsum}  
\usepackage{thmtools}
\usepackage{setspace}
\titleformat{\paragraph}[hang]{\normalfont\normalsize\bfseries}{\theparagraph}{1em}{}
\titlespacing*{\paragraph}{0pt}{3.25ex plus 1ex minus .2ex}{0.5em}
\makeatletter
\DeclareRobustCommand{\change}{%
	\@bsphack
	\leavevmode
	\color{magenta}%
	\@esphack
}
\DeclareRobustCommand{\stopchange}{%
	\@bsphack
	\normalcolor
	\@esphack
}
\makeatother

\stepcounter{secnumdepth}
\stepcounter{tocdepth}

\usepackage{hyperref}
\hypersetup{
  colorlinks   = true, 	
  urlcolor     = blue, 	
  linkcolor    = blue, 	
  citecolor   = magenta 	
}

\title{
	\vskip-1.3cm
Hidden $sl(2)-$Symmetry of the Generalized Landau-Zener Vibronic Model
}

\author{
	L. M. Nieto$^1$\footnote{luismiguel.nieto.calzada@uva.es, ORCID: \href{http://orcid.org/0000-0002-2849-2647}{0000-0002-2849-2647}} \ and 
	S. Zarrinkamar$^{1,2}$\footnote {saber.zarrinkamar@uva.es, ORCID: \href{http://orcid.org/0000-0001-9128-4624}{0000-0001-9128-4624}} 
	\\  [1ex]
	\small
	$^1$\,Departamento de F\'{\i}sica Te\'{o}rica, At\'{o}mica y \'{O}ptica, and Laboratory for Disruptive Interdisciplinary \\ 
				\small
		Science (LaDIS),  Universidad de Valladolid, 47011 Valladolid, Spain\\ [1ex]
	         \small
	$^2$\,Departament of Basic Sciences, Garmsar Branch,
	Islamic Azad University, Garmsar, Iran
}

\begin{document}
	
	\maketitle

\begin{abstract}
The one-dimensional harmonic vibronic model, which is a generalization of the so-called linear Landau-Zener model and appears in the form of coupled Schr\"{o}dinger equations, is revisited. After decoupling the components, the resulting fourth-order equation is shown to have a hidden $sl(2)$ algebra. The so-called exceptional part of the spectrum is then expressed in a rather simple way. For completeness, the eigenfunctions are obtained via the Bethe ansatz approach directly in position space. 
\end{abstract}


\section{Introduction}
The tem {\it vibronic}, which is a combination of the words vibrational and electronic terms, includes spin and nuclear Hamiltonians as well as their interaction \cite {1961, Popkov PRA 2005, Main Paper 0, Main Paper 2009, Main Paper 2010, Main Paper 2012}. 
Vibronic models are present in many areas of physics and their interface with modern technologies, including quantum computing \cite {Science 2005}, quantum optics \cite {Nature Photonics 2015}, quantum cellular automata \cite {Chem 2015 cellular automata}, molecular dimers \cite {Nature 2015 dimer}, chemical physics \cite {Paz Chem 2020}, Rydberg atoms \cite {Schmelcher Rydberg 2023}, molecule magnets \cite {Nature Communication 2024}, and quantum advantage \cite {Nature 2024}. 

However, from the point of view of mathematical physics, this is a difficult problem that cannot be solved simply analytically, even in apparently simple cases.
This is because we have a system of coupled spinor components. In particular, for the simple case of a one-dimensional problem in Cartesian coordinates, which will be considered in the next section, the problem appears in the form of a fourth-order differential equation in position space.
The classification of fourth-order linear differential equations with variable coefficients, which may appear in quantum physics, is something that goes beyond the scope of this article and requires a rather exhaustive analysis. However, in summary, we can say that these problems have not been considered in depth in the analytical field due to the large number of technical problems they present.

In this work we aim to use the powerful tool of Lie symmetries to partially address this problem. The main thrust of our work comes from the novel idea of Zhang \cite{Zhang 2012, Zhang 2013, Zhang Annals 2016}, 
which extends to the fourth-order case the standard idea of quasi-exact solvability in a class of second-order differential equations, originally proposed in the works of Turbiner, Ushveridze, Kamran, Gonzalez-Lopez and Olver \cite {Ushveridze, Turbiner88, Kamran 1990, Artemio94, Turbiner}. 

The structure of the present work is as follows. In Section~\ref{sec2} we first review the main equation of the one-dimensional vibronic model which contains the harmonic oscillator plus a linear interaction \cite {Main Paper 2009} and which can be considered as a generalized Landau-Zener problem. 
Then, using some transformations, the problem becomes a fourth-order differential equation after decoupling the components. 
In Section~\ref{sec3} we then review, in a fairly compact manner, the main idea of quasi-exact solvability for the unfamiliar reader and for the sake of continuity. 
It is then shown that the problem has a hidden structure by which the exceptional part of the exceptional energy can be  derived in a simple, conceptually brief manner and without lengthy calculations. 
In Section~\ref{sec4} we also derive the eigenfunctions of the problem using the Bethe ansatz approach. 
The paper ends in Section~\ref{sec5} with final comments and possible ideas for future research.

\section{Vibronic model}\label{sec2}

In their pioneering work \cite {1961} Fulton and Gouterman proved that the vibronic coupling in molecules and the exciton coupling of dimers can be included in a $2\times 2$ matrix Hamiltonian in the nuclear coordinates. They started from the Hamiltonian
	\begin{equation}\label {hamiltonian}
         \widehat{H}=T_n+T_e+V(q,Q),
	\end{equation}
where, as the indices indicate, $T_n$ and $T_e$ represent the nuclear and electronic kinetic energy, respectively, and $V(q,Q)$ stands for the potential energy which depends on the electronic and nuclear coordinates respectively denoted by $q$ and $Q$. In \eqref {hamiltonian}, some simplifications have been made including neglecting the uniform motion of the center of mass and the rotation of the molecule. It is also obvious that a spin-dependent term is not included. Working on the basis of the adiabatic approximation, the equations in a two-state linear curve crossing model appear in the form (see  \cite {1961} for a detailed derivation)
	\begin{equation}
		\begin{gathered}
			-\frac{\hbar^2}{2m}\frac{d^2\psi_1}{dx^2}+\left(V_{11}(x)-E\right)\psi_1=-V_{12}(x)\psi_2,\\
			-\frac{\hbar^2}{2m}\frac{d^2\psi_2}{dx^2}+\left(V_{22}(x)-E\right)\psi_2=-V_{21}(x)\psi_1,
		\end{gathered}
	\end{equation}
where $m$ is the particle mass, the coordinate $x$ is defined in the full line,  $E$ stands for the energy, and $\psi_1$ and $\psi_2$ are the two components of the wave function. Here, we consider the frequently used model where the two diagonal components are considered to be linear plus harmonic, and the off-diagonal ones are considered to be constants (denoted below by $V$). Therefore, we start from the main equations \cite {Main Paper 2009, Main Paper 2010, Main Paper 2012}
	\begin{equation}\label {Virbronic 1st}
		\begin{gathered}
			-\frac{\hbar^2}{2m}\frac{d^2\psi_1}{dx^2}+\left(\frac{m\Omega^2x^2}{2}-F_1x\right)\psi_1+V\psi_2=E\psi_1,\\
			-\frac{\hbar^2}{2m}\frac{d^2\psi_2}{dx^2}+\left(\frac{m\Omega^2x^2}{2}-F_2x\right)\psi_2+V\psi_1=E\psi_2.
		\end{gathered}
	\end{equation}
The latter, via $x=X+\frac{F_2}{m\Omega^2}$, is rewritten as 
	\begin{equation} \label {Virbronic 2nd}
		\begin{gathered}
-\frac{\hbar^2}{2m}\frac{d^2\psi_1}{dX^2}+\left(\frac{m\Omega^2}{2}X^2+(F_2-F_1)X\right)\psi_1+V\psi_2=\left(E-\frac{F_2^2}{2m\Omega^2}+\frac{F_1F_2}{m\Omega^2}\right)\psi_1,\\
-\frac{\hbar^2}{2m}\frac{d^2\psi_2}{dX^2}+\left(\frac{m\Omega^2}{2}X^2\right)\psi_2+V\psi_1=
\left(E+\frac{F_2^2}{2m\Omega^2}\right)\psi_2.
		\end{gathered}
	\end{equation}
Applying the change of variable
	\begin{equation}\label {change of Variable 2}
		\begin{aligned}
			X=\left(\frac{\hbar}{m\Omega}\right)^{1/2}z,
		\end{aligned}
	\end{equation}
the two equations  \eqref {Virbronic 2nd} become
	\begin{equation} \label{gathered2}
		\begin{gathered}
-\frac{1}{2}\frac{d^2\psi_1}{dz^2}+\left(\frac{1}{2}z^2+(F_2-F_1)\left(\frac{1}{\hbar m \Omega^3}\right)^{1/2}z\right)\psi_1+\frac{V}{\hbar \Omega}\psi_2=\frac{1}{\hbar \Omega}\left(E-\frac{F_2^2}{2m\Omega^2}+\frac{F_1F_2}{m\Omega^2}\right)\psi_1,\\
-\frac{1}{2}\frac{d^2\psi_2}{dz^2}+\frac{1}{2}z^2\psi_2+\frac{V}{\hbar \Omega}\psi_1=
\frac{1}{\hbar \Omega}\left(E+\frac{F_2^2}{2m\Omega^2}\right)\psi_2.
		\end{gathered}
	\end{equation}
Using the tranformations
	\begin{equation}\label {change of Variable}
		\begin{aligned}
                          \psi_{1,2} (z)=e^{-z^2/2}\ y_{1,2} (z), 
		\end{aligned}
	\end{equation}
the two equations  \eqref{gathered2} take the form
	\begin{equation}
		\begin{gathered}
\frac{1}{2}\frac{d^2y_1}{dz^2}-z\frac{dy_1}{dz}-\frac{V}{\hbar \Omega}y_2+\left(\left(F_2-F_1\right)\left(\frac{1}{\hbar m \Omega^3}\right)^{1/2}z+\frac{1}{\hbar \Omega}\left(E-\frac{F_2^2}{2m\Omega^2}+\frac{F_1F_2}{m\Omega^2}\right)-\frac{1}{2}\right)y_1=0,\\
\frac{1}{2}\frac{d^2y_2}{dz^2}-z\frac{dy_2}{dz}- \frac{V}{\hbar \Omega}y_1+\left(\frac{1}{\hbar \Omega}\left(E+\frac{F_2^2}{2m\Omega^2}\right)-\frac{1}{2}\right)y_2=0.
		\end{gathered}
	\end{equation}
We can decouple these two equations by solving for $y_2$ in the first one and replacing it in the second, so that we arrive at
\begin{eqnarray} \label {before trans}
\widehat{H}_4\, y_1\!\!&\!=\!&\!\! \frac{1}{4}\frac{d^4y_1}{dz^4}-z\frac{d^3y_1}{dz^3}+
\left[z^2+\frac{F}{2}z+ \left(\frac{E_1}{2}+\frac{E_2}{2}-1\right) \right]\frac{d^2y_1}{dz^2} \nonumber   \\
&&  \!\!  +\left[-Fz^2+z (1-E_2-E_1)\right]\frac{dy_1}{dz}+\left[FE_2z+(E_1E_2-v^2)\right]y_1=0,
\label {differential form}
	\end{eqnarray}
where the explicit form of the fourth-order differential operator $\widehat{H}_4$, which can be considered a kind of ``Hamiltonian'' for this problem, is clear, and where we have used the abbreviations
\begin{equation} \label {shorthand}
		\begin{gathered}
F=\left(F_2-F_1\right)\left(\frac{1}{\hbar m \Omega^3}\right)^{1/2},\qquad \qquad
E_1=\frac{1}{\hbar \Omega}\left(E-\frac{F_2^2}{2m\Omega^2}+\frac{F_1F_2}{m\Omega^2}\right)-\frac{1}{2},\\
E_2=\frac{1}{\hbar \Omega}\left(E+\frac{F_2^2}{2m\Omega^2}\right)-\frac{1}{2},\qquad\qquad\qquad
v=\frac{V}{\hbar \Omega}.
		\end{gathered}
\end{equation}
Eq. \eqref {differential form} is an ordinary fourth-order differential equation with variable coefficients. To our best knowledge, this equation has not been reported as a generalized special function of mathematical physics, nor has been recognized as a named equation. 
It should be noted that papers on fourth-order differential equations in physics are really scarce, although there are important cases of such forms. One might think that the problem can be solved simply by numerical techniques, but no one can deny the merits of analytical techniques, including their deep insight into the physics of the system as well as their pedagogical importance. We can comment on the need for analytical approaches for a variety of other reasons, but this is beyond the scope of the present paper and we hope to return to it in a subsequent study.

\section{The hidden $sl(2)$ symmetry and the spectrum}\label{sec3}

To obtain the energy spectrum of the system, let us first review the basic idea of the Lie algebraic approach, or quasi-exact solvability, in a summarized manner. The term quasi-exact solvability comes from the fact that only part of the spectrum, and not all states, can be derived in this way. Let us first review the idea as it is used for second-order differential equations and then generalize it to the fourth-order case.

\subsection {Lie algebraic approach  for second-order differential equations}
The most general form of the second-order quasi-exactly solvable differential operator which can be expressed as an $sl(2)$ algebra is \cite {Ushveridze, Turbiner88, Artemio94, Turbiner}
\begin{equation}
\widehat{H}_{QES}=C_{++}{J}_n^+{J}_n^++C_{+0}{J}_n^+{J}_n^0+C_{+-}{J}_n^+{J}_n^-+  C_{0-}{J}_n^0{J}_n^-+C_{--}{J}_n^-{J}_n^-+C_+{J}_n^++C_0{J}_n^0+C_-{J}_n^-+C,
\end{equation}
in which
\begin{equation}\label{GenerJJJ}
		J_n^+ =z^2\,\frac{d}{dz}-n\,z ,\qquad
		J_n^0 = z\,\frac{d}{dz}-\frac n2,\qquad
		J_n^- = \frac{d}{dz},
\end{equation}
where the generators satisfy
\begin{equation}
[J^+_n,J^-_n]=-2J^0_n,  \ \ \ \   [J^{\pm}_n,J^0_n]=\mp J^{\pm}_n.
\end{equation}
As a result, it can be shown that $\widehat{H}_{QES}$ preserves the finite-dimensional space of polynomials of the form 
\begin{equation}
	y_n (z)=\sum_{m=0} ^n{c_m z^m}.
\end{equation}
Using \eqref{GenerJJJ}, the operator $\widehat{H}_{QES}$ can be written as
\begin{equation} \label {2nd hamiltonial}
\widehat{H}_{QES}=P_4(z)\frac{d^2}{dz^2}+P_3(z)\frac{d}{dz}+P_2(z),
\end{equation}
with
\begin{equation}\label{dieciocho}
P_4(z)=\sum_{k=0} ^4a_kz^k, \qquad P_3(z)=\sum_{k=0} ^3b_kz^k, \qquad P_2(z)=\sum_{k=0} ^2c_kz^k,
\end{equation}
and more precisely
\begin{equation}\label{Coefficients}
	\begin{aligned}
	&	 P_4(z)=C_{++}z^4+C_{+0}z^3+C_{+-}z^2+C_{0-}z+C_{--},\\
	&	P_3(z)=C_{++}(2-2n)z^3+\left (C_++C_{+0}\left (1-\frac{3n}{2}\right)\right)z^2+\left (C_0-nC_{+-}\right)z+\left (C_--\frac{n}{2}C_{0-} \right), \\
	&	 P_2(z)=C_{++}n(n-1)z^2+\left(\frac{n^2}{2}C_{+0}-nC_+\right)z+\left (C-\frac{n}{2}C_0 \right).
	\end{aligned}
\end{equation}

\subsection{Hidden $sl(2)$ symmetry for fourth-order differential equations}

Based on the novel idea of Zhang \cite {Zhang 2013, Zhang Annals 2016}, the problem has a hidden $sl(2)$-algebraization if and only if 
\begin{equation} \label {condition}
b_3=-2(n-1)a_4, \qquad c_2=n(n-1)a_4, \qquad c_1=-n\left[(n-1)a_3+b_2\right].
\end{equation}
Now, following the identity \cite {Zhang Annals 2016}
\begin{equation} \label{Identity} 
z\frac{d^3}{dz^3}=J^0\left(J^-\right)^2+\frac{n}{2}\frac{d^2}{dz^2},
\end{equation}
the ``Hamiltonian'' $\widehat{H}_4$ in \eqref {differential form} can be written in the form
	\begin{equation} \label {Hamiltonian}
		\begin{gathered}
\widehat{H}_4\,  y_1=\left[ \frac{1}{4}(J^-)^4-J^0\left(J^-\right)^2\right] y_1
+\left[z^2+\frac{F}{2}z+ \left(-\frac{n}{2}+\frac{E_1}{2}+\frac{E_2}{2}-1\right) \right]\frac{d^2y_1}{dz^2}\\
\qquad\quad +\left[-Fz^2+z\left(1-E_2-E_1\right)\right]\frac{dy_1}{dz}
+\left[FE_2z+(E_1E_2-v^2)\right]y_1=0,
		\end{gathered}
	\end{equation}
where we have intentionally kept the equivalent second order part of $\widehat{H}_4\,  y_1$ in differential form to comment on the hidden symmetry of the system. It is seen that the $n$-dependent term in \eqref {Identity} has affected the second order part of the fourth-order differential equation. The energy relation from the third equation in \eqref {condition}, i.e. $c_1=-n\left[(n-1)a_3+b_2\right]=-nb_2$, is found as $E_2=n$, or, from \eqref{shorthand}, more explicitly in the form
\begin{equation} \label {energy 1}
\frac{1}{\hbar \Omega}\left(E+\frac{F_2^2}{2m\Omega^2}\right)-\frac{1}{2}=n.
\end{equation}
To make a comparison with \cite {Main Paper 2009}, noting that in their notation 
\begin{equation}
b=\frac{F_2}{(m\hbar \Omega^3)^{1/2}}, \qquad  \epsilon=\frac{E}{\hbar \Omega}+\frac{1}{2},
\end{equation}
Eq. \eqref {energy 1} can be written as
\begin{equation}
\frac{E}{\hbar \Omega}+\frac{1}{2}=\epsilon=(n+1)-\frac{b^2}{2}=k-\frac{b^2}{2},
\end{equation}
which is the same result obtained in \cite {Main Paper 2009} and possesses the familiar linear form of the harmonic oscillator in terms of the quantum number. Also, doing the shift transformation on the upper equation in \eqref {Virbronic 1st}, one may obtain the other half of the quasi-exact spectrum with the same behavior of the other component.

\section{The Bethe ansatz approach and  eigenfunctions: a generalization to the fourth-order case}\label{sec4}

Although the main purpose of the present work was demonstrating the hidden $sl(2)$-symmetry of the vibronic model in its fourth-order form and directly in position space, by which the exceptional part of the spectrum was reported, it is interesting to comment on the eigenfunctions as well. To this aim, we make use of the Bethe ansatz method similar to what is done in \cite {Zhang 2012, Zhang 2013}. The idea in a nutshell is that an ansatz solution is proposed to the equation, and after manipulations, a meromorphic equation is obtained. Based on comparison of different powers on both sides of the arising equation, required material to obtain the spectrum and the eigenfunctions of the system is obtained. There are two main references which are strongly recommended for the unfamiliar reader before starting this section. The first is \cite {Zhang 2012} in which the idea is applied to second-order ordinary differential equations and the basic formalism is obtained. The second is \cite {Zhang 2013} which generalizes the primary idea to a fourth-order ordinary differential equation,  the two-photon and two-mode Rabi model in Bargmann space in this case. The generalization to the fourth-order case can be done, rather straightforwardly, using proper series identities and careful use of residue calculus.

Let us first review the basic idea from \cite {Zhang 2012} which works for Fuchsian-type equatios. If we propose a solution of the form
\begin{equation} \label{ansatz}
y(z)=\prod_{i=1}^n(z-z_i), \quad n=1,2,\dots , 
	\end{equation}
for the second-order Hamiltonian \eqref  {2nd hamiltonial}, it can be found that  \cite {Zhang 2012}
\begin{eqnarray}\label {General Form}
&& 
\hskip-1.7cm
-c_0-\sum_{i=1}^n {\rm Res} 
\left[  \frac{\widehat{H}_{QES}\, y}{y} , z=z_i \right] 
  =\left[n(n-1)a_4+nb_3+c_2\right]z^2  \nonumber  \\
                        &&   
                           +\left[\left(2(n-1)a_4+b_3\right)\sum_{i=1}^nz_i+n(n-1)a_3+nb_2+c_1\right]z  \nonumber  \\
                        &&   +\left(2(n-1)a_4+b_3\right)\sum_{i=1}^nz_i^2+2a_4 \sum_{i<j}^nz_iz_j
                           +\left(2(n-1)a_3+b_2\right)\sum_{i=1}^nz_i+n(n-1)a_2+nb_1,
	\end{eqnarray}
where ${\rm Res}[f(z),z=z_i]$ indicates the residue of $f(z)$ at $z=z_i$. In fact, we have a constant term on the left-hand side, and a meromorphic function on the right-hand side, from which the required data to report the solution is provided. In more precise words, the residue term, which must be vanishing, gives the $z_i$. The coefficients of $z^2$ and $z$ on the right-hand side normally determine the spectrum of the system and both should vanish to assure that the equality holds. The remaining constant terms on both sides, can be interpreted as a restriction among the parameters, which is the price ought to be paid to obtain an analytical solution. Using \cite {Zhang 2013}, which has to be read to carefully to understand the details of calculations, the idea is generalized to the fourth-order case.

Having reviewed the basic idea of the approach, we now propose a solution of the form \eqref {ansatz} for $y_1(z)$ in \eqref {before trans}, which yields for the quotient $\frac1{y_1}\widehat{H}_4\, y_1$:
	\begin{eqnarray}
0\!\!&\!=\!&\!\!    E_1E_2 -v^2 +\frac{1}{4} \sum_{i=1}^n\frac{1}{z-z_i} \sum_{p\neq l \neq j\neq i}^n \frac{4}{(z_i-z_p)(z_i-z_l)(z_i-z_j)}
-z \sum_{i=1}^n\frac{1}{z-z_i} \sum_{ l \neq j\neq i}^n \frac{3}{(z_i-z_l)(z_i-z_j)} \nonumber \\
&& \!\!  +\left[z^2+\frac{F}{2}z+\left(\frac{E_1}{2}+\frac{E_2}{2}-1\right)\right] \sum_{i=1}^n\frac{1}{z-z_i} \sum_{ j\neq i}^n \frac{2}{(z_i-z_j)} \nonumber \\
&& 
\!\!   +\left[-Fz^2+z\left (1-E_2-E_1\right)\right]\sum_{i=1}^n\frac{1}{z-z_i}+FE_2z. 
 \label {subs}
	\end{eqnarray}
From here we can easily evaluate  the following residues 
	\begin{eqnarray} 
	{\rm Res} 
\left[  \frac{\widehat{H}_4\, y_1}{y_1} , z=z_i \right] 
 \!\!&\!=\!&\!\!
\frac{1}{4} \sum_{p\neq l \neq j\neq i}^n \frac{4}{(z_i-z_p)(z_i-z_l)(z_i-z_j)}
 -z_i\sum_{ l \neq j\neq i}^n \frac{3}{(z_i-z_l)(z_i-z_j)}  \label {Residues}
 \\
&&\!\!+\left[z_i^2+\frac{F}{2}z_i+\left(\frac{E_1}{2}+\frac{E_2}{2}-1\right)\right] \sum_{ j\neq i}^n \frac{2}{(z_i-z_j)}
+\left[-Fz_i^2+z_i\left (1-E_2-E_1\right)\right],  \nonumber
	\end{eqnarray}
which determines the $z_i$ in the ansatz proposed in \eqref {ansatz}. For example, for the first state, i.e. $z_i=z_1$, we have $-Fz_i^2+z_i\left (1-E_2-E_1\right)=0$ as all other terms vanish (here $E_{1,2}$ are in fact linear functions of the exceptional energy $E_n$ as indicated in \eqref {shorthand} and thus vary for different $n$). A systematic process can be done for higher states. 
However, as can be seen, the higher the state, the more complicated the calculations. Nevertheless, the approach is quite systematic and can be carried out. 

Let us now go back to the arising meromorphic equation \eqref {General Form}. In fact, this equation can be used here in the fourth-order case recalling that the effect of third- and fourth-order terms are already included in the $z_i$ and the residue at each $z_i$. As a result, we may conclude that in our case, i.e. for \eqref {before trans}:
\begin{itemize}
\item 
The first term on the right-hand side of \eqref{General Form}, i.e. the coefficient of $z^2$, yields nothing since  $a_4=b_3=c_2=0$ as follows by comparing \eqref {before trans} with \eqref{dieciocho}. 

\item
The second term on the right-hand side of \eqref{General Form} gives $nb_2+c_1=0$, i.e. the coefficient of $z$, yields the exceptional energy already obtained in \eqref {energy 1}. This is because $a_3=0$ as follows by comparing \eqref {before trans} with \eqref{dieciocho}.
\item
Taking into account that, comparing \eqref {before trans} with \eqref{dieciocho}, we also get that $a_2=1$, $b_1=1-E_2-E_1$ 
and $b_2=-F$, the constant terms on both sides of of \eqref{General Form} yield
	\begin{equation}\label {restrict}
		\begin{gathered}
			-c_0= v^2-E_1E_2=-F\sum_{i=1}^nz_i+n(n-1)+n\left(1-E_2-E_1\right),
		\end{gathered}
	\end{equation}
which can be considered as a restriction among the parameters. More explicit form of this restriction can be easily obtained using the value of $z_i$, from \eqref {Residues}, and $E_{1,2}$ from \eqref {shorthand} and \eqref {energy 1} for each level/state.

\end{itemize}

\section{ Conclusion}\label{sec5}

A two-state vibronic model was considered in one spatial dimension with a harmonic plus linear interaction for the diagonal scattering matrix element and constant terms for the off-diagonal ones. We worked directly in the position space instead of transforming the problem into other spaces or using integral transforms. It was shown that the problem, via proper transformations and identities, has a hidden $sl(2)$ symmetry which gives the exceptional part of the spectrum in a simple manner. This is of particular interest due to various complexities and challenges of the other approaches. 
In particular, the present problem, which of great importance from both research and  pedagogical points of view, was solved by first transforming the equation into $p-$representation space and then working on the arising Heun-like equation in the transformed space \cite{Main Paper 2009}.
 Here, however, the exceptional part of the spectrum was derived from quite basic ideas of quasi-exactly solvable models and elements of group theory. 

Although the main purpose of this work was recognizing and introducing the hidden symmetry of the problem, the eigenfunctions were also obtained via the Bethe  ansatz approach in a systematic manner. The latter has its own limitations and complexity and is certainly rather lengthy. Nevertheless, when compared to the other parallel approaches, it might still look more economical. 

The idea might be generalized to other classes of vibronic models, on which we are working. We hope the present work renews the interest in the analytical structure of the other vibronic and two- and/or multi-level models, which have not been extensively explored for many realistic terms mainly due to their higher-order structure. It is also an interesting survey to consider the idea in different fields and in particular within the so-called Generalized Uncertainty Principle (GUP) formalism where we have to deal with higher-order wave equations, and in particular, the fourth and sixth order Sch{\"o}dinger equations for a frequently used form of GUP originating from the interface of quantum physics and gravity. Work in this direction is also in progress.

\section*{Acknowledgements}
This research was supported by the Q-CAYLE project, funded by the European Union-Next Generation UE/MICIU/Plan de Recuperacion, Transformacion y Resiliencia/Junta de Castilla y Leon (PRTRC17.11), and also by project PID2023-148409NB-I00, funded by MICIU/AEI/10.13039/501100011033. Financial support of the Department of Education of the Junta de Castilla y Leon and FEDER Funds is also gratefully acknowledged (Reference: CLU-2023-1-05).

	\end{document}